\documentclass[twocolumn,aps,superscriptaddress,showpacs,showkeys,amsmath,amssymb,floatfix, longbibliography,nofootinbib,prx]{revtex4-1}

\usepackage{threeparttable}
\usepackage{xurl}
\usepackage[colorlinks=true,citecolor=blue,filecolor=blue,linkcolor=blue,urlcolor=blue,pdftex,breaklinks=true]{hyperref}
\usepackage{currfile} 
\usepackage{graphicx}
\usepackage{dcolumn}
\usepackage{bm}
\usepackage[section]{placeins} 

\usepackage{isotope} 
\usepackage{upgreek} 
\usepackage{cancel}
\usepackage{soul} 
\usepackage[separate-uncertainty=true]{siunitx} 
\usepackage[normalem]{ulem}
\usepackage{color}
\usepackage{amsmath}
\usepackage{multirow}
\usepackage{extarrows}
\usepackage{xspace}
\usepackage{orcidlink} 
\usepackage[mathlines]{lineno}
\newcommand{\decay}[2]{\xLongrightarrow[\mathrm{{#2}}]{\alpha\mathrm{{#1}}}}
\newcommand{\decayb}[1]{\xlongrightarrow[\mathrm{{#1}}]{\upbeta}}
\newcommand{\decaybBi}[1]{\xlongrightarrow[\mathrm{{#1}}]{\upbeta\,(64\%\,\mathrm{BR})}}

\newcommand*\patchAmsMathEnvironmentForLineno[1]{%
  \expandafter\let\csname old#1\expandafter\endcsname\csname #1\endcsname
  \expandafter\let\csname oldend#1\expandafter\endcsname\csname end#1\endcsname
  \renewenvironment{#1}%
     {\linenomath\csname old#1\endcsname}%
     {\csname oldend#1\endcsname\endlinenomath}}%
\newcommand*\patchBothAmsMathEnvironmentsForLineno[1]{%
  \patchAmsMathEnvironmentForLineno{#1}%
  \patchAmsMathEnvironmentForLineno{#1*}}%
\AtBeginDocument{%
\patchBothAmsMathEnvironmentsForLineno{equation}%
\patchBothAmsMathEnvironmentsForLineno{align}%
\patchBothAmsMathEnvironmentsForLineno{flalign}%
\patchBothAmsMathEnvironmentsForLineno{alignat}%
\patchBothAmsMathEnvironmentsForLineno{gather}%
\patchBothAmsMathEnvironmentsForLineno{multline}%
}

\newcommand\exposure{$2.46\,\mathrm{t} \cdot \mathrm{y}$}
\newcommand\neutrinoenergylow{\qty{17}{keV}} 

\newcommand\solarppfluxresultnominalcombined{$(10.2 \pm 2.0) \times 10^{10}$\,cm$^{-2}$s$^{-1}$}
\newcommand\solarnusignificancnominalcombined{$5.0\,\sigma$}
\newcommand\solarppdeviationnominalcombined{$1.9\,\sigma$}
\newcommand\combinedpvalue{0.43}

\newcommand\solarppfluxresultcarboncombined{$(8.7^{+2.8}_{-2.1}) \times 10^{10}$\,cm$^{-2}$s$^{-1}$}
\newcommand\solarnusignificanccarbon{$4.3\,\sigma$}
\newcommand\solarppdeviationcarboncombined{$1.2\,\sigma$}

\newcommand\solarppsignificancrrpa{$5.0\,\sigma$}
\newcommand\solarppfluxresultrrpacombined{$(12.7 \pm 2.5) \times 10^{10}$ cm$^{-2}$s$^{-1}$}
\newcommand\dechalflife{$(8.4 \pm 0.9) \times 10^{21}\,\mathrm{y}$}
\newcommand\decdeviation{$1.7\,\sigma$}

\newcommand\solarppratioraw{\qty{89.0}{\percent}}

\begin{document}


\title{First Measurement of Solar Neutrinos through Elastic Neutrino-Electron Scattering at the keV Scale}





\newcommand{\bologna}{\affiliation{Department of Physics and Astronomy, University of Bologna and INFN-Bologna, 40126 Bologna, Italy}}
\newcommand{\chicago}{\affiliation{Department of Physics, Enrico Fermi Institute \& Kavli Institute for Cosmological Physics, University of Chicago, Chicago, IL 60637, USA}}
\newcommand{\coimbra}{\affiliation{LIBPhys, Department of Physics, University of Coimbra, 3004-516 Coimbra, Portugal}}
\newcommand{\columbia}{\affiliation{Physics Department, Columbia University, New York, NY 10027, USA}}
\newcommand{\lngs}{\affiliation{INFN-Laboratori Nazionali del Gran Sasso and Gran Sasso Science Institute, 67100 L'Aquila, Italy}}
\newcommand{\mainz}{\affiliation{Institut f\"ur Physik \& Exzellenzcluster PRISMA$^{+}$, Johannes Gutenberg-Universit\"at Mainz, 55099 Mainz, Germany}}
\newcommand{\mpik}{\affiliation{Max-Planck-Institut f\"ur Kernphysik, 69117 Heidelberg, Germany}}
\newcommand{\munster}{\affiliation{Institut f\"ur Kernphysik, University of M\"unster, 48149 M\"unster, Germany}}
\newcommand{\nikhef}{\affiliation{Nikhef and the University of Amsterdam, Science Park, 1098XG Amsterdam, Netherlands}}
\newcommand{\nyuad}{\affiliation{New York University Abu Dhabi - Center for Astro, Particle and Planetary Physics, Abu Dhabi, United Arab Emirates}}
\newcommand{\purdue}{\affiliation{Department of Physics and Astronomy, Purdue University, West Lafayette, IN 47907, USA}}
\newcommand{\rice}{\affiliation{Department of Physics and Astronomy, Rice University, Houston, TX 77005, USA}}
\newcommand{\stockholm}{\affiliation{Oskar Klein Centre, Department of Physics, Stockholm University, AlbaNova, Stockholm SE-10691, Sweden}}
\newcommand{\subatech}{\affiliation{SUBATECH, IMT Atlantique, CNRS/IN2P3, Nantes Universit\'e, Nantes 44307, France}}
\newcommand{\torino}{\affiliation{INAF-Astrophysical Observatory of Torino, Department of Physics, University  of  Torino and  INFN-Torino,  10125  Torino,  Italy}}
\newcommand{\ucsd}{\affiliation{Department of Physics, University of California San Diego, La Jolla, CA 92093, USA}}
\newcommand{\wis}{\affiliation{Department of Particle Physics and Astrophysics, Weizmann Institute of Science, Rehovot 7610001, Israel}}
\newcommand{\zurich}{\affiliation{Physik-Institut, University of Z\"urich, 8057  Z\"urich, Switzerland}}
\newcommand{\paris}{\affiliation{LPNHE, Sorbonne Universit\'{e}, CNRS/IN2P3, 75005 Paris, France}}
\newcommand{\freiburg}{\affiliation{Physikalisches Institut, Universit\"at Freiburg, 79104 Freiburg, Germany}}
\newcommand{\napels}{\affiliation{Department of Physics ``Ettore Pancini'', University of Napoli and INFN-Napoli, 80126 Napoli, Italy}}
\newcommand{\nagoya}{\affiliation{Kobayashi-Maskawa Institute for the Origin of Particles and the Universe, and Institute for Space-Earth Environmental Research, Nagoya University, Furo-cho, Chikusa-ku, Nagoya, Aichi 464-8602, Japan}}
\newcommand{\laquila}{\affiliation{Department of Physics and Chemistry, University of L'Aquila, 67100 L'Aquila, Italy}}
\newcommand{\tokyo}{\affiliation{Kamioka Observatory, Institute for Cosmic Ray Research, and Kavli Institute for the Physics and Mathematics of the Universe (WPI), University of Tokyo, Higashi-Mozumi, Kamioka, Hida, Gifu 506-1205, Japan}}
\newcommand{\kobe}{\affiliation{Department of Physics, Kobe University, Kobe, Hyogo 657-8501, Japan}}
\newcommand{\kit}{\affiliation{Institute for Astroparticle Physics \& Institute of Experimental Particle Physics, Karlsruhe Institute of Technology, 76021 Karlsruhe, Germany}}
\newcommand{\tsinghua}{\affiliation{Department of Physics \& Center for High Energy Physics, Tsinghua University, Beijing 100084, P.R. China}}
\newcommand{\ferrara}{\affiliation{INFN-Ferrara and Dip. di Fisica e Scienze della Terra, Universit\`a di Ferrara, 44122 Ferrara, Italy}}
\newcommand{\groningen}{\affiliation{Nikhef and the University of Groningen, Van Swinderen Institute, 9747AG Groningen, Netherlands}}
\newcommand{\westlake}{\affiliation{Department of Physics, School of Science, Westlake University, Hangzhou 310030, P.R. China}}
\newcommand{\shenzhen}{\affiliation{School of Science and Engineering, The Chinese University of Hong Kong (Shenzhen), Shenzhen, Guangdong, 518172, P.R. China}}
\newcommand{\coimbrapoli}{\affiliation{Coimbra Polytechnic - ISEC, 3030-199 Coimbra, Portugal}}
\newcommand{\heidelberg}{\affiliation{Kirchhoff-Institute for Physics, Heidelberg University, 69120 Heidelberg, Germany}}
\newcommand{\bucknell}{\affiliation{Department of Physics \& Astronomy, Bucknell University, Lewisburg, PA, USA}}
\newcommand{\isct}{\affiliation{Department of Physics, School of Science, Institute of Science Tokyo, Meguro, Tokyo, 152-8551, Japan}}





\author{E.~Aprile\,\orcidlink{0000-0001-6595-7098}}\columbia
\author{J.~Aalbers\,\orcidlink{0000-0003-0030-0030}}\groningen
\author{K.~Abe\,\orcidlink{0009-0000-9620-788X}}\tokyo
\author{M.~Abu~Rmilah\,\orcidlink{0009-0007-9750-6655}}\wis
\author{M.~Adrover\,\orcidlink{0123-4567-8901-2345}}\zurich
\author{S.~Ahmed~Maouloud\,\orcidlink{0000-0002-0844-4576}}\paris
\author{L.~Althueser\,\orcidlink{0000-0002-5468-4298}}\munster
\author{B.~Andrieu\,\orcidlink{0009-0002-6485-4163}}\paris
\author{E.~Angelino\,\orcidlink{0000-0002-6695-4355}}\lngs\chicago
\author{D.~Ant\'on~Martin\,\orcidlink{0000-0001-7725-5552}}\chicago
% ...
\collaboration{XENON Collaboration\textsuperscript{**}}\noaffiliation








\date{\today}

\begin{abstract}
    \noindent 
We report on the first measurement of low-energy solar neutrinos through elastic neutrino-electron scattering in a dark matter experiment, establishing the lowest energy threshold for any neutrino detection to date.
The measurement utilizes data from the first two science runs of XENONnT, corresponding to an exposure of {\exposure}, and covers electron recoil energies between \qty{1}{keV} and \qty{140}{keV}, providing sensitivity to solar neutrinos with energies down to {\neutrinoenergylow}. 
We reject the background-only hypothesis with a statistical significance of {\solarnusignificancnominalcombined} and measure a solar $pp$ neutrino flux of {\solarppfluxresultnominalcombined}. The result is larger, but statistically consistent with the previous measurement by Borexino at {\solarppdeviationnominalcombined}.
Together with recent observations of coherent elastic neutrino--nucleus scattering of \isotope[8]{B} solar neutrinos in XENONnT and other liquid-xenon time projection chambers, these results demonstrate the growing potential of liquid xenon detectors for neutrino physics down to the keV-scale and represent an important milestone towards a next-generation multipurpose observatory.
\end{abstract}


{
\let\clearpage\relax
\maketitle
}
\def\thefootnote{*}\footnotetext{\email{kobayashi.masatoshi@isee.nagoya-u.ac.jp}}
\def\thefootnote{\textdagger}\footnotetext{\email{ylin3@uni-muenster.de}}
\def\thefootnote{\textdaggerdbl}\footnotetext{\email{yuema@physics.ucsd.edu}}
\def\thefootnote{\S}\footnotetext{\email{dwenz@uni-muenster.de}}
\def\thefootnote{\P}\footnotetext{\email{florian.joerg@physik.uzh.ch}}
\def\thefootnote{**}\footnotetext{\email{xenon@lngs.infn.it}}
\def\thefootnote{\textdagger\textdagger}\footnotetext{Also at Coimbra Polytechnic - ISEC, 3030-199 Coimbra, Portugal}

\section{Introduction}
\label{sec:introduction}
\begin{figure*}[t]
	\centering
	\includegraphics[width=0.95\textwidth]{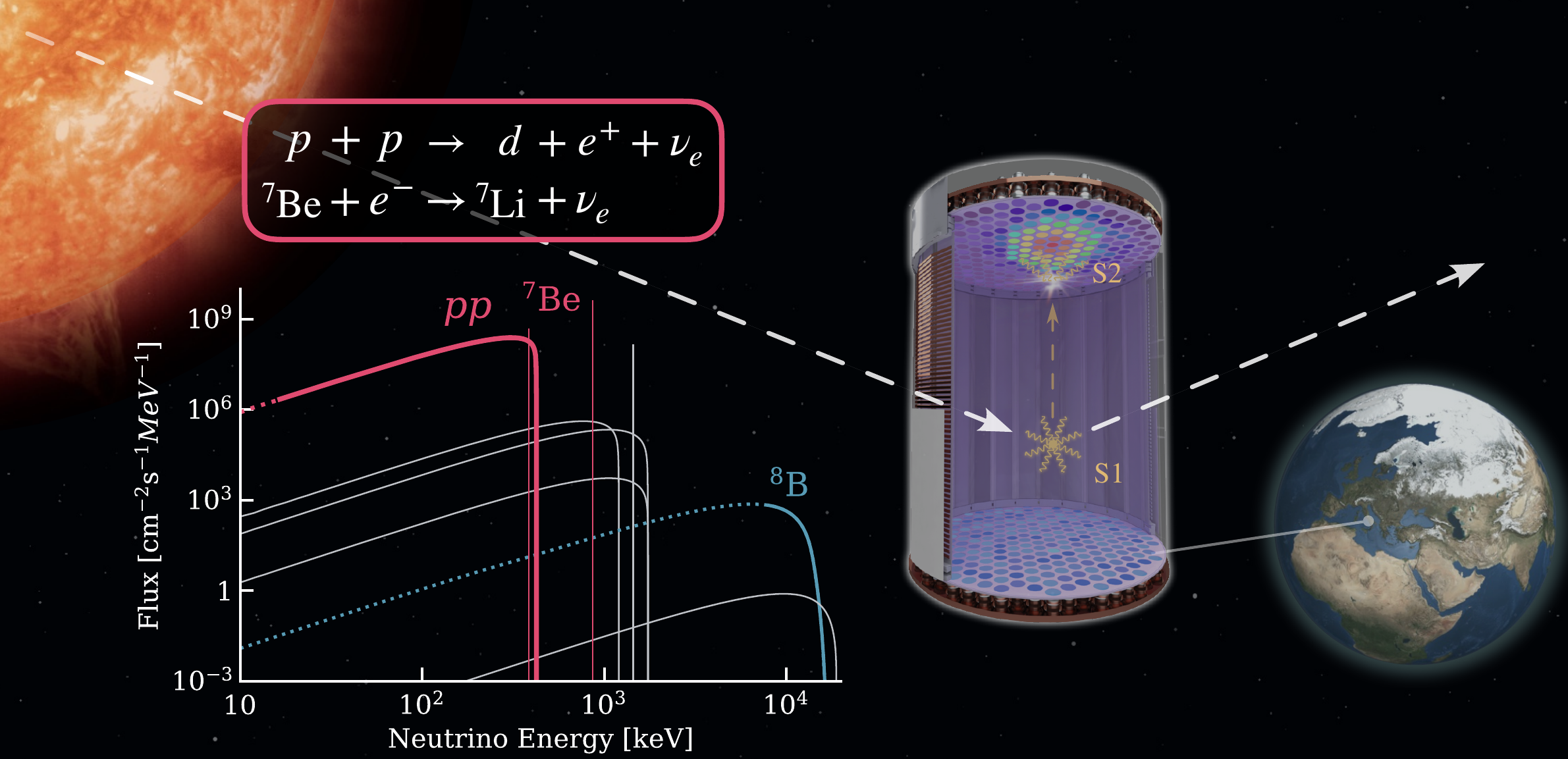}
	\caption{
    Solar neutrinos are produced in the Sun’s core via a chain of different nuclear processes. The box in the upper left illustrates the two most dominant neutrino sources considered in this study. Solar $pp$ neutrinos are produced via the fusion of two protons into a deuteron, a positron, and an electron neutrino. The $^7$Be neutrinos are emitted during the electron-capture decay of $^7$Be to $^7$Li.
    The produced neutrinos propagate from the Sun's core to Earth, where they have a small chance to be detected through the simultaneous measurement of two scintillation signals (S1 and S2) inside the liquid xenon (LXe) time projection chamber (TPC) of XENONnT (see Sect. \ref{sec:detector_data_observables}), which is located at the INFN Laboratori Nazionali del Gran Sasso underground laboratory in Italy.
    The different solar neutrino flux components at Earth as a function of neutrino energy are shown in the bottom left figure~\cite{Bahcall:1997eg,otherSolarNv}. 
    So far only $^8$B neutrinos above an energy of about $8\,\mathrm{MeV}$ have been successfully measured in LXe TPCs~\cite{XENONnT_B8_SR2, XENONnT_B8_SR1, PandaX_B8, LZ_B8} (solid cyan line), while in this work we report a measurement of solar neutrinos at energies down to {\neutrinoenergylow} (solid red lines). Dashed lines indicate the fraction of the solar neutrino spectrum below the detector threshold of the respective analysis. Solar neutrinos from other processes that contribute subdominantly to the presented study are indicated in gray. Continuous spectra are reported in units of $[\mathrm{cm}^{-2} \mathrm{s}^{-1} \mathrm{MeV}^{-1}]$, while line sources are in units of $[\mathrm{cm}^{-2} \mathrm{s}^{-1}]$.
    Images adapted from the following sources: Sun~\cite{ESA_SOHO_Sun}, Earth~\cite{NASA_BlueMarble}, and TPC~\cite{radonRemoval_level}.} 
	\label{Fig:solar_neutrino_flux}
\end{figure*}

\noindent
Precision measurements of solar neutrinos are a cornerstone of modern solar and particle physics. Produced in nuclear reactions that power the Sun, solar neutrinos provide direct information about the physical conditions in the solar core~\cite{bahcall1989neutrino, Adelberger2011}. Their flux comprises several distinct components, conventionally identified by their production mechanism, including $pp$, $^7$Be, $pep$, $^8$B, and CNO neutrinos~\cite{SSM, Adelberger2011}. Their energy spectra are illustrated in the lower-left inset of Fig.~\ref{Fig:solar_neutrino_flux}. 
Because of their extremely small interaction cross section $\mathcal{O}(10^{-45}\,\mathrm{cm}^2)$, most solar neutrinos escape the Sun essentially unimpeded, undergoing only oscillations between electron, muon, and tau flavor eigenstates. This makes them a powerful probe of the Standard Solar Model (SSM)~\cite{SSM,Haxton2013}.
Moreover, solar neutrinos provide an additional sensitive probe of other neutrino properties, like the neutrino magnetic moment and non-standard interactions, thereby enabling together with flavor oscillations precision tests of the Standard Model (SM) and searches for physics beyond it\,\cite{XENONnT_SR0_lowER,PDG2024}. 

Despite their abundance, the tiny interaction cross section of solar neutrinos makes their detection challenging. Since the pioneering detection by the Homestake experiment in 1968~\cite{Homestake}, solar neutrinos and their properties have been primarily studied by experiments utilizing tens of tonnes of Gallium (SAGE~\cite{SAGE}, GALLEX/GNO~\cite{GALLEX_GNO}), or hundreds of tonnes of liquid scintillator (Borexino~\cite{Borexino}, KamLAND~\cite{KamLAND_B8}), water (Super-Kamiokande~\cite{SuperK}) or heavy water (SNO experiment~\cite{SNO}) as target material, whose large target masses compensate for the extremely small neutrino interaction probability. 
This changed recently, with the first measurement of coherent elastic neutrino-nucleus scattering (CE$\nu$NS) induced by solar $^8$B neutrinos in tonne-scale dual-phase xenon time projection chambers (TPCs)~\cite{XENONnT_B8_SR2, XENONnT_B8_SR1, PandaX_B8, LZ_B8}. Since liquid xenon (LXe) TPCs have originally been designed for the search for Dark Matter (DM) in the form of Weakly Interacting Massive Particles (WIMPs), they provide a low detection threshold ($\mathcal{O}(1\,\mathrm{keV})$), excellent particle discrimination, and an exceptionally low intrinsic radioactive background~\cite{XENONnT_SR0_lowER}. This makes LXe TPCs attractive for a wide range of rare-event searches, including studies of solar neutrino properties~\cite{XLZD}. 

Although $^8$B neutrinos only account for less than \qty{0.01}{\percent} in the solar neutrino spectrum~\cite{otherSolarNv}, their high energies make them generally more accessible than other solar neutrinos. 
In particular, solar $pp$ neutrinos that dominate the solar neutrino spectrum with about \qty{90}{\percent} are challenging to detect given their low energy of less than $425\,\mathrm{keV}$~\cite{Bahcall:1997eg}. 
This places their CE$\nu$NS signal below the detection threshold of current LXe TPCs, making elastic neutrino-electron scattering (ES) the dominant detection channel. 
Here we report on the first detection of such low-energy solar neutrinos via neutrino-electron scattering in XENONnT, and the subsequent measurement of the solar $pp$ neutrino flux. This measurement facilitates the detection of neutrinos at the lowest energies ever observed ($\lesssim100\,\mathrm{keV}$). It was enabled by recent advances in reducing the dominant background component from $^{214}$Pb, a daughter isotope of \isotope[222]{Rn}~\cite{radonRemoval_level}, together with dedicated calibrations and background estimation campaigns that constrain the associated systematic uncertainties.

\section{Detector, Data and Observables}
\label{sec:detector_data_observables}
\noindent

\subsection{Experiment}
\noindent

The XENONnT experiment is installed at the INFN Laboratori Nazionali del Gran Sasso underground laboratory in Italy, beneath \qty{1400}{\meter} of rock overburden (\qty{3800}{\meter} water equivalent), providing a factor $\mathcal{O}(10^6)$ reduction in the flux of cosmic muons~\cite{XENONnT_Instrumental}. The main detector of the experiment is a dual-phase LXe TPC with a \qty{5.9}{tonne} LXe active target mass in an approximately cylindrical volume. The target is viewed by two inward-facing arrays of light sensors located at the bottom and top (see Fig.~\ref{Fig:solar_neutrino_flux}), which are composed of 494 photomultiplier tubes (PMTs)~\cite{XENONnT_Instrumental,XENONnT_PMT}. Particle interactions in LXe produce scintillation photons and ionization electrons as detectable quanta. The scintillation photons are recorded directly by the PMTs as a prompt scintillation signal (S1). The ionization electrons drift under an electric field to a thin gaseous xenon layer, where they are extracted by a stronger field, producing a secondary proportional scintillation signal (S2) that is recorded by the PMT arrays~\cite{XENONnT_Instrumental,XENONnT_SignalReconstruction}.
Prompt and secondary scintillation signals are reconstructed from individual PMT signals using XENONnT's open source reconstruction framework~\cite{XENONnT_SignalReconstruction,straxen}. 

The combined measurement of S1 and S2 signals provides the basis for the background discrimination in LXe TPCs. The ratio of S1 to S2 encodes information about the type of interaction, distinguishing electronic recoils (ER) with the xenon shell electrons from nuclear recoils (NR)~\cite{XENONnT_SR1_WIMP}. 
While ER/NR discrimination is essential for the search of DM and CE$\nu$NS, lower-energy solar neutrinos are predominantly detected via elastic neutrino--electron scattering, rendering such discrimination for the dominant $\beta$ and $\gamma$-ray backgrounds inapplicable. 
Thus, external backgrounds are primarily suppressed through a three-dimensional position reconstruction achieved by the combined S1 and S2 readout (see Appendix~\ref{sec:appendix_fdv}).
The spatial resolution in XENONnT is approximately \qty{5}{mm} ~\cite{XENONnT_SignalReconstruction}. Combined with the high density of LXe ($\rho_\mathrm{LXe}\approx\qty{2.9}{\gram \per \cubic \centi \meter}$), this enables the definition of an inner fiducial volume that is effectively shielded against $\beta$s and $\gamma$ rays emitted by detector materials. Together with a comprehensive radioassay campaign and two outer veto detectors housed in a \qty{700}{tonne} water tank, this renders external backgrounds from surrounding materials sub-dominant in the solar neutrino search~\cite{RadioPurity,NeutronVeto}.

The main challenge for the solar neutrino search arises from intrinsic radioactive impurities uniformly dissolved in the LXe volume. Natural xenon consists primarily of stable isotopes, with the exceptions of the very long-lived \isotope[124]{Xe} and \isotope[136]{Xe}, which decay via second-order weak transition~\cite{XENON1T_DEC,XENONnT_SR0_lowER}. More problematic for the solar neutrino search are the $\beta$-emitting anthropogenic \isotope[85]{Kr} and the \isotope[222]{Rn} daughter \isotope[214]{Pb}. The recoil spectra of the emitted $\beta$s from the two isotopes exhibit a similar spectral resemblance to solar neutrinos and dominate the low-energy ER background in XENONnT. To reduce these backgrounds, the experiment utilizes dedicated cryogenic distillation systems~\cite{kryptonDistillation,radonRemoval}. The krypton distillation system has demonstrated the capability to reduce the $^{\mathrm{nat}}$Kr concentration to $\lesssim$\qty{38}{ppq} (parts-per-quadrillion mol/mol) (\cite{kryptonDistillation2017}, updated according to Sec.~\ref{subsec:kr85}), rendering the \isotope[85]{Kr} background less important. Thanks to the higher vapor pressure of argon compared to krypton, the cosmogenically produced isotope \isotope[39]{Ar} is even more efficiently removed~\cite{argonDistillation}, making this background completely negligible in XENONnT (see Appendix~\ref{app:ar39_background}). Consequently, the dominant background remaining in the solar neutrino analysis is \isotope[222]{Rn}, whose specific activity was reduced for the first time by using a dedicated radon removal system to about \qty{0.9}{\micro \becquerel \per \kilo \gram}~\cite{radonRemoval_level}, which is equivalent to the expected solar neutrino interaction rate in XENONnT. 

\subsection{Data and signal efficiency}
\label{sec:data_efficiency}

\noindent
This analysis uses data recorded during the first (SR0) and second (SR1) science runs of XENONnT, lasting from May 2021 to November 2021 and from May 2022 to August 2023, respectively. For SR0, the livetime-corrected exposure is \qty{96.5}{d}. Due to an elevated background level in the first part of SR1~(SR1a), only the second half of the science run (SR1b), corresponding to an exposure of \qty{117.8}{d}, is used (see Sec.~\ref{subsec:kr85} for more details). Detector conditions, including pressure and temperature, were monitored throughout both science runs and found to be stable within \qty{1.0}{\percent} and \qty{0.2}{\percent} respectively~\cite{XENONnT_SR1_WIMP, XENONnT_B8_SR1}. The signal reconstruction efficiency for S1 and S2 signals was monitored by regular $^{83\mathrm{m}}\mathrm{Kr}$ calibrations~\cite{XENONnT_SignalReconstruction}.
Spatial and temporal variations in the light and charge collection efficiencies for S1 and S2 signals were corrected, while the remaining fluctuations were propagated as a systematic uncertainty of approximately \qty{3}{\percent} on the reconstructed energy~\cite{XENONnT_SignalReconstruction, XENONnT_SR1_WIMP}. 

\begin{figure}[t]
	\centering
	\includegraphics[width=\columnwidth]{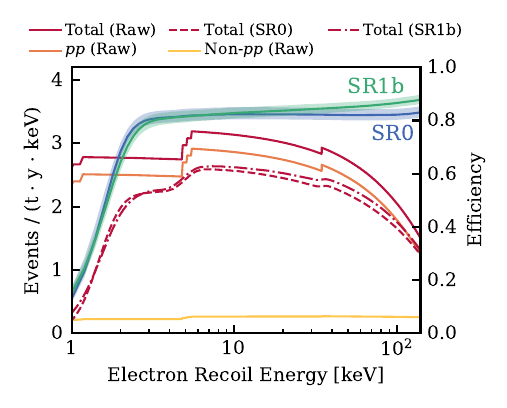}
	\caption{Solar neutrino spectrum as a function of recoil energy before (solid red line) and after (red dashed lines) applying the signal selection efficiency together with detector energy resolution and bias. The total efficiencies for SR0 and SR1b (right axis) are shown in blue and green, respectively. The solar $pp$ neutrino signal contributes with {\solarppratioraw} (orange) to the total solar neutrino spectrum inside the ROI. Other contributions from $^7$Be, $^8$B, $^{13}$N, $^{15}$O, $^{17}$F, $hep$ and $pep$ (yellow) are subdominant.
    }
	\label{Fig:2_sr01_solarpp_eff_signal_log}
\end{figure}

The reconstructed energy of an electron recoil is given by ${E_\mathrm{rec}\,=\,W\times(\mathrm{cS1}/g_1+\mathrm{cS2}/g_2)}$, and is the only observable used in the statistical inference.
The recoil energy is proportional to the corrected peak areas cS1 and cS2, and the parameter $W=\qty{13.7\pm0.2}{eV}$~\cite{DahlThesis}, which is the average energy required to produce a single scintillation photon or ionization electron. The photon and electron gains, $g_1$ and $g_2$, are detector-specific constants that represent the average peak area measured by the PMT arrays in number of photoelectrons (PE) for a single photon or electron produced. 
They are calibrated independently for each science run using several monoenergetic calibration lines ~\cite{XENONnT_SignalReconstruction,XENONnT_SR1_WIMP}. The gains were found to be $g_1=\qty{0.151\pm0.001}{PE/photon}$ and $g_2=\qty{16.5\pm0.6}{PE/electron}$ for SR0 and $g_1=\qty{0.137\pm0.001}{PE/photon}$ and $g_2=\qty{16.9\pm0.5}{PE/electron}$ for SR1. To propagate the energy resolution and bias of the detector to the expected signal spectrum, the monoenergetic calibration lines are fitted in $E_\mathrm{rec}$ for both science runs~\cite{XENONnT_SignalReconstruction}. The impact of the energy resolution on the signal spectrum is shown in Fig.~\ref{Fig:2_sr01_solarpp_eff_signal_log}. 
The uncertainties in the energy resolution and bias are propagated as systematic uncertainties in the final statistical inference. 

The region of interest (ROI) for this analysis is defined between a recoil energy of $[1,\,140]\,\si{\kilo\electronvolt}$, maximizing the sensitivity to solar neutrinos, while avoiding higher-energy spectral features, such as multiple-scattering $\gamma$-rays and a prominent line of $^{131\mathrm{m}}\mathrm{Xe}$ from neutron activation following calibrations. Additional quality selections and limited signal reconstruction efficiency close to the detector threshold at about \qty{1}{keV} reduce the total reconstruction efficiency of the solar neutrino signal. The signal efficiency was evaluated using XENONnT's open source waveform simulation framework~\cite{fuse} in combination with \isotope[220]{Rn} calibration data~\cite{XENONnT_SignalReconstruction}. The total efficiency is shown in Fig.~\ref{Fig:2_sr01_solarpp_eff_signal_log} and is approximately \qty{80\pm2}{\percent} across most of the ROI. Details of the efficiency determination are provided in Appendix~\ref{sec:appendix_efficiency}.

Equally important is a precise knowledge of the fiducial mass, as any bias in its estimate directly leads to a bias in the solar neutrino flux estimate. The fiducial volume boundaries were optimized in radius and height $(R^2, Z)$ using a signal-to-background figure of merit.
The geometrical size of the fiducial volume was corrected for distortions in the position reconstruction using $^{83\mathrm{m}}\mathrm{Kr}$ calibration data and waveform simulations~\cite{XENONnT:2023field}. Its uncertainty was conservatively estimated to be \qty{3}{\percent}~\cite{ToschiThesis}. The LXe density is determined from the monitored temperature and pressure within each science run. The final fiducial mass is estimated to be $(4.13 \pm 0.12)$\,\si{tonne} in SR0 and $(4.24 \pm 0.13)$\,\si{tonne} in SR1. Details of the fiducial-volume optimization are provided in Appendix~\ref{sec:appendix_fdv}.

\section{Signal and background}
\label{sec:signal_background}
\noindent
\subsection{Solar neutrino signal}
\label{sec:solarpp_signal}

The differential rate ${\mathrm{d}R_i}/{\mathrm{d}E_r}$ as a function of recoil energy $E_r$ for neutrino-electron scattering under the free-electron approximation (FEA) can be described as~\cite{FEAsignalModel}:
\begin{equation}
	\frac{\mathrm{d}R_i}{\mathrm{d}E_r} = \sum_j \int{P_{ej}}\frac{\mathrm{d}\Phi_i}{\mathrm{d}E_\nu}\frac{\mathrm{d}\sigma_j}{\mathrm{d}E_r}\mathrm{d}E_\nu,
\label{Eqn:diffRate}
\end{equation}
where $E_\nu$ is the neutrino kinetic energy, $i$ the neutrino component from each process, such as $pp$ and $^7$Be, $j$ the neutrino flavor (electron $e$, muon $\mu$ or tau $\tau$), $P_{ej}$ the oscillation probability from flavor $e$ to $j$ \cite{SurvivalProb}, and $\mathrm{d}\Phi/\mathrm{d}E_\nu$ the differential neutrino flux at Earth (see Fig.~\ref{Fig:solar_neutrino_flux}). The ES channel is sensitive to electron neutrinos via both charged-current and neutral-current interactions, while it is sensitive to muon and tau neutrinos only via neutral currents. In natural units, $\hbar=c=1$, the differential cross section $\mathrm{d}\sigma/\mathrm{d}E_r$ can be expanded~\cite{FEAsignalModel}:
\begin{equation}
	\frac{\mathrm{d}\sigma}{\mathrm{d}E_r} = \frac{2G_F^2m_e}{\pi} \left[ g_L^2 + g_R^2 \left(1-\frac{E_r}{E_\nu} \right)^2 -g_Lg_R\frac{m_eE_r}{E_\nu^2}\right], 
    \label{Eqn:xsec}
\end{equation}
where the coupling constants are $g_R = \sin^2{\theta_\mathrm{W}}$, $g_L = \sin^2{\theta_\mathrm{W}} + 1/2$ for $\nu_e$ and $g_L = \sin^2{\theta_\mathrm{W}} - 1/2$ for $\nu_{\mu,\tau}$ given the Weinberg angle $\theta_\mathrm{W}$. 

In reality, electrons in xenon are bound in atomic shells. Therefore, at low recoil energies, the interaction is forbidden if the deposited energy is insufficient to release the electrons from their shell. This effect can be accounted for to some extent using the stepping approximation (FEA-stepping model)~\cite{RRPAsignalModel}:
\begin{equation}
	\frac{d\sigma}{dE_r} = \sum^Z_{s=1} \theta(E_r-E_b^{(s)})\frac{d\sigma^{(s)}}{dE_r}, 
\end{equation}
where $Z$ is the atomic number. The contribution from the $s$-th electron is weighted by a step function determined by its binding energy, $E_b^{(s)}$.

The solar neutrino detection rate in LXe as a function of deposited energy is given in Fig.~\ref{Fig:2_sr01_solarpp_eff_signal_log}. Solar $pp$ neutrinos dominate the signal within the ROI. The next largest contribution originates from $^7$Be neutrinos with approximately \qty{10}{\percent}, while the remaining ($^8$B, $^{13}$N, $^{15}$O, $^{17}$F, $hep$ and $pep$) contributes less than \qty{1}{\percent}. The fraction for each flux component is fixed according to the theoretical high-metallicity SSM (B16-GS98)~\cite{otherSolarNv}, giving a solar $pp$ component fraction of \solarppratioraw~within the ROI. The total solar neutrino spectrum is scaled by a single rate parameter (see Sec.~\ref{sec:StatisticalInference}), as neutrinos from different channels cannot be distinguished in this study. The total signal rate $R_i$ is calculated by integrating Eq.~(\ref{Eqn:diffRate}) and accounting for the detector response described in Sec.~\ref{sec:data_efficiency}. Given the electron-recoil energy threshold of $1~\si{keV}$, the minimum neutrino energy detectable by the XENONnT detector is approximately {\neutrinoenergylow}.

A refined model using the relativistic random phase approximation (RRPA) that accounts for atomic effects, such as relativistic corrections, has also been developed in Ref.~\cite{RRPAsignalModel}. However, the FEA-stepping model is adopted as the primary signal model in this work because the current calculations of the RRPA model are available only up to $E \leq 32$\,keV and therefore do not cover the full ROI of this analysis.
A more detailed discussion of this RRPA model and corresponding results is provided in Appendix~\ref{app:rrpa}. 

\subsection{Backgrounds}
\label{sec:constrainingTheBackgrounds}

The primary challenge of this analysis is the spectral degeneracy between the solar neutrino signal and the dominant backgrounds, whose spectra are nearly flat across the ROI~\cite{XENONnT_SR0_lowER}. Consequently, the following discussion is organized according to the importance of each background component. We first discuss the dominant \isotope[214]{Pb} background together with \isotope[212]{Pb}, followed by the smaller but spectrally similar contributions from \isotope[85]{Kr} and material-induced $\gamma$ rays. Then, a tritium-like component is considered. Although its spectral shape differs from that of the solar neutrino signal, its low endpoint energy places most of its spectrum in the region with the highest signal-to-background ratio. Finally, we discuss the \isotope[136]{Xe} two-neutrino double-beta decay background, whose spectral shape is distinct from that of the signal, but it is among the largest backgrounds within the ROI. The remaining background components are less relevant and therefore summarized in a final section, with additional details provided in Appendix~\ref{app:details_background_eval}.

\subsubsection*{Lead-214 and lead-212}
\label{subsec:radonAnalysis}

The dominant background comes from $\beta$ decays of $^{214}$Pb. It is part of the decay chain of the radioactive noble gas \isotope[222]{Rn}, which is constantly emanated from trace impurities of uranium within detector materials and distributes homogeneously throughout the LXe volume. The decay chain following \isotope[222]{Rn}~\cite{ESNDF} is given by

\begin{eqnarray}
&\dots~ \isotope[222]{Rn} \decay{}{3.8\,d} \isotope[218]{Po} \decay{}{3.1\,min} \isotope[214]{Pb}\notag\\
&\decayb{27\,min} \isotope[214]{Bi} \decayb{20\,min} \isotope[214]{Po} \decay{}{160\,\upmu s} \isotope[210]{Pb}~\dots.
\label{eq:rn222_chain}
\end{eqnarray}

The activity of \isotope[214]{Pb} can be constrained using the more prominent $\alpha$ events, whose decay energies can be fully captured within the xenon TPC (see Appendix~\ref{app:details_background_eval}).
However, due to convection, drift, and the plate-out process, charged radon progeny can perturb the equilibrium of the decay chain, leading to smaller activities of the daughter isotopes and thereby introducing a significant systematic uncertainty~\cite{radonRemoval_level, LZ:2025xxf}.
While this conveniently suppresses the $^{214}$Pb background, constraining the systematic requires a dedicated calibration to precisely determine the corresponding activity ratio between the rate of $^{214}$Pb $\beta$-decays and that of $^{222}$Rn $\alpha$-decays.
The calibration was performed during SR1b\,\cite{XENONnT_Bi214Spec}, using a \isotope[222]{Rn} source with an emanation rate of approximately \SI{2}{Bq}\,\cite{Jorg:2022tli}. 
\begin{figure}[t]
    \hspace*{-0.5cm}
	\raggedright
	\includegraphics[width=\columnwidth]{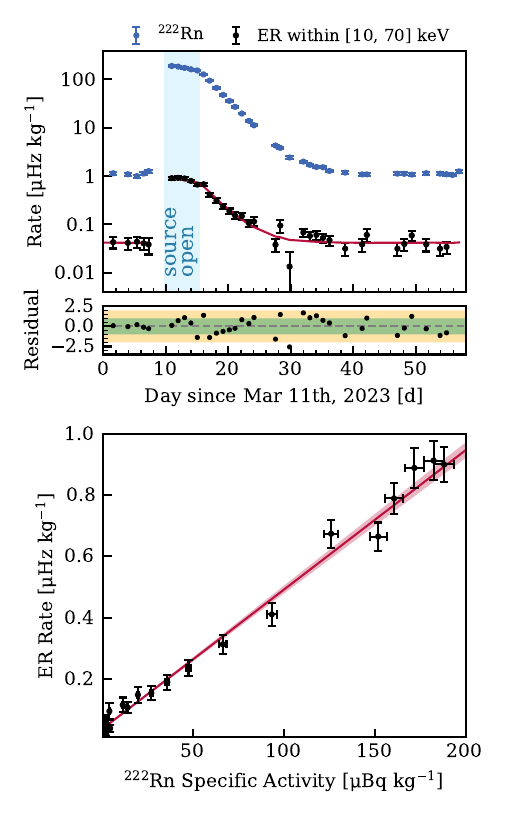}
	\caption{
    $^{214}$Pb/\isotope[222]{Rn} activity ratio using \isotope[222]{Rn} calibration and adjacent science data. The top panel shows the time evolution of the $^{222}$Rn $\alpha$ activity (blue) and ER event rate within $[10,\,70]\,\si{keV}$ (black). The source injection is indicated by the cyan shaded period. The bottom panel shows the correlation between the two rates, along with a linear fit and its uncertainty band (red). Based on this fit and the \isotope[222]{Rn} $\alpha$ rate, a prediction for the ER rate is given as the red curve in the top plot, whose residual is derived in comparison to the measured ER event rate (black).
    }
	\label{Fig:rateRatioFit_Rn222calibration}
\end{figure}
The top panel of Fig.~\ref{Fig:rateRatioFit_Rn222calibration} shows the time evolution of the activity during calibration.
During this time, the event rate within the ER ROI is dominated by $^{214}$Pb $\beta$ decays. 
The bottom panel of Fig.~\ref{Fig:rateRatioFit_Rn222calibration} shows the correspondence between the \isotope[222]{Rn} $\alpha$ and the ER event rate within a narrower window of $[10,\,70]\,\si{keV}$ to exclude the influence of \isotope[133]{Xe} produced in a preceding neutron calibration. 
The correlation between the ER event rate and \isotope[222]{Rn} rate is then obtained through a linear fit. The offset of the linear fit that describes all non-\isotope[214]{Pb} contributions pivots on science data before and after the calibration. During this period, the activity ratio, $^{214}$Pb/$^{222}$Rn, is evaluated as $0.66 \pm 0.03$ based on the slope obtained from the fit and accounting for the detector effects as well as the branching ratio of the decay to the ground state \cite{XENONnT_PbBR}. The systematic uncertainty from the spectral difference between forbidden and allowed transitions is considered at the inference stage (see Sec.~\ref{sec:StatisticalInference}, Table~\ref{Tab:postInferenceSystematics}).

Although the specific activity of \isotope[220]{Rn} is approximately one order of magnitude smaller compared to \isotope[222]{Rn}, the $^{212}$Pb $\beta$ decay constitutes another important background following the decay chain,
\begin{eqnarray}
&\dots~ \isotope[220]{Rn} \decay{}{56\,s} \isotope[216]{Po} \decay{}{150\,ms} \isotope[212]{Pb}\notag\\
&\decayb{11\,h} \isotope[212]{Bi} \decaybBi{60\,min} \isotope[212]{Po} \decay{}{300 \,ns} \isotope[208]{Pb},
\label{eq:rn220_chain}
\end{eqnarray}
with the remaining \qty{36}{\percent} branching ratio (BR) of \isotope[212]{Bi} undergoing alpha decays. Similarly, a \isotope[220]{Rn} calibration was performed to estimate the $^{212}$Pb rate using the $\alpha$ events. Given the much shorter half-life of \isotope[220]{Rn} and \isotope[216]{Po}, we rely only on the $\alpha$ decays of \isotope[212]{Bi} and \isotope[212]{Po} to constrain its activity (see Appendix~\ref{app:details_background_eval}).

The resulting predictions for the specific activities of \isotope[214]{Pb} and \isotope[212]{Pb} for both science runs are shown in Table~\ref{Tab:RnResults}. The difference in \isotope[214]{Pb} originates from two different operation modes of the radon removal system for the two science runs~\cite{radonRemoval_level}. While during SR0 only a small gaseous xenon flow of about \qty{20}{slpm} was extracted and purified, the operation mode in SR1 was changed such that an additional \qty{200}{slpm} liquid flow was redirected through the radon removal system. This increases the reduction in radon from a factor of about two in SR0 to a factor of about four in SR1. Due to the very short lifetime of \isotope[220]{Rn}, the activity of \isotope[212]{Pb} is not affected by this operation~\cite{radonRemoval_level}.

\begin{table}[tb]
    \setlength{\extrarowheight}{2pt}
	\centering
    \caption{Estimation of the specific activity of $^{214}$Pb and $^{212}$Pb. The $^{214}$Pb activity uncertainty includes a correlated uncertainty between SR0 and SR1b, accounting for \qty{4.1}{\percent} of the activity value.}
	\begin{tabular}{c|cc}
		\hline
        \hline
		& SR0 & SR1b \\
        \hline
        $^{214}$Pb activity (\qty{}{\micro \becquerel \per \kilo \gram}) & 1.39 $\pm$ 0.07 & 0.76 $\pm$ 0.03 \\
        $^{212}$Pb activity (\qty{}{\micro \becquerel \per \kilo \gram}) & 0.038 $\pm$ 0.007 & 0.034 $\pm$ 0.004 \\
        \hline
        \hline
	\end{tabular}
	\label{Tab:RnResults}
\end{table}

\subsubsection*{Krypton-85}
\label{subsec:kr85}

Following its separation from ambient air during production, xenon retains trace amounts of krypton, including the anthropogenic isotope $^{85}$Kr. Its relatively high $Q_\beta$-value of \SI{687}{\kilo\electronvolt} yields a $\beta$ decay spectrum that is nearly flat at low energies. This makes it an important background component in the search for solar neutrino signals. The $^{85}$Kr concentration in the detector is estimated from the concentration of natural krypton in xenon, $^\mathrm{nat}$Kr/Xe, and the isotopic abundance, $^{85}$Kr/$^\mathrm{nat}$Kr.

The concentration of natural krypton is primarily constrained by the measurement of samples taken from the detector through Rare Gas Mass Spectrometry~(RGMS)~\cite{RGMS,autoRGMS}. 
A dedicated calibration campaign was carried out to improve the overall RGMS measurements used in this study. 
Measurements using multiple new calibration standards with a $^{\mathrm{nat}}$Kr concentration precisely known at the percent level showed a consistently \SI{46}{\percent} higher $^{\mathrm{nat}}$Kr content compared to former measurements.  
The adjustment has been applied throughout the analysis. 

\begin{figure}
	\centering
	\includegraphics[width=\columnwidth]{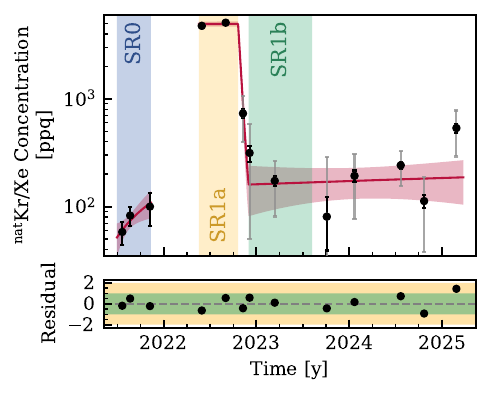}
	\caption{The $^\mathrm{nat}\mathrm{Kr}$ concentration in xenon as a function of time. The RGMS measurements are shown as the black data points. The SR1a period (orange) is followed by a krypton distillation campaign and reaches the low krypton SR1b period (green). For SR1, the solid line (red) indicates the fit function incorporating two different models. Gray error bars indicate the additional systematic uncertainty based on the $\chi^2$ scaling discussed in the text. The SR0 (blue) fit is also shown~\cite{XENONnT_SR0_lowER}. The shades (red) indicate the fit uncertainties. The residuals are calculated based on the scaled uncertainties. Only data from SR0 and SR1b are used as science data.}
\label{Fig:5_251118_combkil_sr1b_fit_slope_fullsr2}
\end{figure}

The $^{\mathrm{nat}}$Kr concentration in XENONnT is monitored through regular RGMS measurements shown in Fig.~\ref{Fig:5_251118_combkil_sr1b_fit_slope_fullsr2}.
The SR0 result is the same as in Ref.~\cite{XENONnT_SR0_lowER}, and SR1 is subdivided into two periods labeled as SR1a and SR1b.
An unintended gas introduction during SR1a increased the $^{\mathrm{nat}}$Kr concentration to approximately \qty{5000}{ppq}, leading to the exclusion of SR1a data for this analysis.
The subsequent removal of krypton via cryogenic distillation is modeled by an empirical exponential decay~\cite{kryptonDistillation}.
For the post-distillation period that includes SR1b, two distinct evolution models are considered: one incorporating a constant outgassing term potentially from internal PTFE components and another assuming zero outgassing. The outgassing rate is constrained using the SR0 values as an informative prior, and the difference between the two evolution models is added as systematic uncertainty~\cite{XENONnT_SR0_lowER}. Furthermore, to account for the instrument-related RGMS systematics since SR1 that caused the overall spread of individual RGMS measurements, an additional systematic uncertainty is introduced by a common scaling factor such that the reduced chi-square is unity in the least-squares fit. This instrument-related uncertainty dominates the systematic uncertainty in SR1. The time-averaged $^{\mathrm{nat}}$Kr concentrations for both SR0 and SR1b are shown in Table~\ref{Tab:KrResults}.

To estimate the final $^{85}$Kr background rate, the $^{\mathrm{nat}}$Kr result must be scaled according to the isotopic abundance of $^{85}$Kr. At the time of installation of XENONnT, the ambient air had an estimated isotopic abundance of $(2.33 \pm 0.14) \times 10^{-11}$\,mol/mol~\cite{autoRGMS}. During SR0, it is assumed that the isotopic abundance changes purely over time through the decay of $^{85}$Kr and is therefore modeled via an exponential decay.
For SR1a, the initial isotopic abundance of $^{85}$Kr is directly measured from the $^{85}$Kr $\beta$ decay rate as it dominates the low-energy background during this period.
By associating the measured rate with the RGMS measurement, the ratio is evaluated as $(2.69 \pm 0.06) \times 10^{-11}$ mol/mol. 
For SR1b, we again consider two scenarios: the outgassing of ambient atmospheric krypton from SR0 versus potentially remnant krypton from SR1a. The median is taken as the nominal value, with half of the difference as the uncertainty. The resulting $^{85}$Kr/$^{\mathrm{nat}}$Kr ratios are summarized in Table~\ref{Tab:KrResults}.

As an independent consistency check, the $^{85}$Kr abundance was validated through a sub-dominant decay branch of $^{85}$Kr (\SI{0.435}{\percent} \cite{NuDataKr85}), which populates the $9/2^+$ metastable state of $^{85}$Rb that de-excites with a half-life of $T_{1/2} = \SI{1.015}{\micro\second}$. This decay enables the search for a unique $\beta$-$\gamma$ delayed coincidence signature and has the strong advantage of constraining the $^{85}$Kr activity directly inside the detector. 
Two complementary methodologies were developed to identify this signature: a cut-based analysis comparing data to simulations within a \SIrange{1.5}{5.0}{\micro\second} delay window between consecutive S1 signals~\cite{YoshinoThesis}, and a machine-learning-based classifier optimized for the shorter \SIrange{0.05}{1.28}{\micro\second} range \cite{HenningThesis}. Although the sensitivity of these analyses is statistically limited by the low branching ratio, the results for the SR1a period are consistent with the primary measurements within $2 \sigma$. Furthermore, for the lower-concentration SR0 and SR1b periods, both studies established upper limits that are fully compatible with the RGMS-derived values.

\begin{table}[tb]
    \setlength{\extrarowheight}{2pt}
	\centering
    \caption{Estimation of the krypton content during the different science runs.}
	\begin{tabular}{c|ccc}
		\hline
        \hline
		& SR0 & SR1a & SR1b \\
        \hline
        $^{\mathrm{nat}}$Kr (ppq) & 82 $\pm$ 23 & 4950 $\pm$ 270 & 170 $\pm$ 70 \\
        $^{85}$Kr/$^{\mathrm{nat}}$Kr ($10^{-11}$) & 2.12 $\pm$ 0.13 & - & 2.3 $\pm$ 0.4 \\
        \hline
        \hline
	\end{tabular}
	\label{Tab:KrResults}
\end{table}

\subsubsection*{Material-induced $\gamma$ background}

All detector materials used in XENONnT were screened and selected for radio-purity~\cite{RadioPurity}. De-excitation $\gamma$ rays emitted by trace amounts of residual radioactive impurities, such as U, Th, K, and Co, are further suppressed by selecting the inner fiducial volume, although a small fraction of these $\gamma$ rays can still reach this volume and deposit energy through Compton scattering. 
While this component is subdominant, these interactions produce an approximately flat electron-recoil spectrum and therefore constitute a background component resembling ES of solar neutrinos.
To this end, each $\gamma$-emitting isotope is simulated using its screened activity with updates from Ref.~\cite{XENONnT_SR1_WIMP}, and the resulting waveforms are processed with the same reconstruction chain as the data, thereby accounting for detector response and reconstruction effects~\cite{XENONnT_SR1_WIMP}. The simulation is normalized by a data-driven material-background sideband, defined by a \SI{2}{cm} thick radial shell immediately outside the fiducial volume after subtracting the contribution of uniformly distributed intrinsic backgrounds estimated from the innermost \SI{1}{tonne} of the fiducial volume. The final material-background model is taken from the high-statistics simulation and normalized using these measured scale factors of $1.47\pm0.11$ for SR0 and $1.04\pm0.12$ for SR1. 
The improved prediction for SR1 compared to SR0 originates from a better understanding of distortions in the electrical drift field and a resulting refined position reconstruction. To account for the discrepancy between the simulation model and the data sideband, the half difference between the two is assigned as a systematic uncertainty. This effect dominates the total uncertainty in SR0, which is \qty{25}{\percent}, while SR1 is dominated by statistical uncertainties with a total uncertainty of \qty{12}{\percent}.

\subsubsection*{Tritium-like background}

During a search for WIMP dark matter utilizing SR1b science data, an excess in the event rate near the detector threshold was observed~\cite{XENONnT_SR1_WIMP}. After a systematic evaluation of detector effects, signal corrections, and event selection criteria, we determined that this feature is not an analysis artifact. Although its continuous shape resembles a $\beta$-like spectrum, the precise identification of its origin remains uncertain due to its absence in SR0 \cite{XENONnT_SR0_lowER}. Its distinct spectral shape makes this additional background component less dominant in the overall background than the flat-shaped components discussed in previous sections. However, it overlaps with the solar neutrino signal below \qty{20}{keV}, where the signal-to-background ratio is highest, increasing its importance relative to other backgrounds with distinct spectral shapes.

To model this component, we use the spectral shape of tritium, a radioactive isotope of hydrogen that undergoes a $\beta$ decay with a $Q_\beta$-value of 18.6~\si{keV}. Tritium is a known background in other LXe TPC experiments~\cite{PandaX:2024cic, LZ_ER2025}, which utilize titrated methane as a calibration source. However, this is not the case for XENONnT and no plausible mechanism for potential tritium contamination was identified, as gaseous xenon passes through a purification system containing a hot zirconium metal getter as well as dedicated hydrogen removal units~\cite{XENONnT_Instrumental}. In the absence of an alternative explanation, we estimated the nominal value for this $^{3}$H-like component by utilizing a sideband region located \SI{1}{\cm} outside the fiducial volume boundary. The volume corresponds to a mass of \SI{0.21}{tonne} and is sufficient to get a statistically relevant estimate, while being small enough to limit the influence from material backgrounds. In this region, the $^{3}$H-like component is fitted together with the remaining flat low-ER background above \SI{20}{keV}. Given the uncertainty surrounding this background component, the rate is set as a free parameter in the inference to be conservative.

\subsubsection*{Xenon-136}
\label{subsec:xenonIsotopes}

Natural xenon itself only contains the two very long-lived isotopes \isotope[124]{Xe} and \isotope[136]{Xe}, which both decay via second-order weak decays~\cite{XENON1T_DEC, XENONnT_SR0_lowER}. For the solar neutrino search, the two-neutrino double-$\beta$ decay of $^{136}$Xe provides the most significant background. 
While its spectrum has a distinct shape, it dominates parts of the ROI, and its half-life of $(2.17 \pm 0.06) \times 10^{21}\,\mathrm{y}$~\cite{EXO200} makes it nearly constant in time, elevating its importance over other xenon isotopes that have either a more distinct spectral shape or are strongly localized in time, e.g, activated isotopes following calibrations with external neutrons. Its rate is constrained by its half-life and the measured isotopic abundance of $(8.97 \pm 0.16) \times 10^{-2}$~\cite{XENON1T_DEC_first, XENON1T_DEC, XENONnT_SR0_lowER, LZ_DEC, PANDAX_DEC}. For the spectral shape,  the Higher-State Dominance (HSD)~\cite{Xe136_HSD_and_Xe134} and Single-State Dominance (SSD)~\cite{Xe136_SSD} theoretical models are considered as shown later in Table~\ref{Tab:postInferenceSystematics}.

\subsubsection*{Other backgrounds}
\label{subsubsec:other_backgrounds}

In addition to the dominant backgrounds discussed above, the background model includes several other components whose spectral shapes are distinct from the solar neutrino signal. Contributions from neutron-activated xenon isotopes like \isotope[133]{Xe} are covered in Appendix~\ref{app:neutronActivation}. 
The accidental coincidence background due to the random pairing of uncorrelated S1 and S2 signals is also included, but its contribution is found to be negligible. Its rate is constrained using the methods described in Appendix~\ref{app:EventReconstructionEfficiency}.
While runs directly following $^{83\mathrm{m}}$Kr calibrations are excluded from the science data, residual $^{83\mathrm{m}}$Kr can still leak into the science data selection. $^{83\mathrm{m}}$Kr has a short half-life of \SI{1.83}{h} and decays via two mono-energetic transitions. It therefore has a very distinct shape and is very localized in time and is thus only included as a free parameter in the fit.
Finally, we also performed a rate measurement of $^{124}$Xe double-electron capture (DEC), treating it as a signal in the inference as presented in Appendix~\ref{app:dec}.

\noindent
\section{Statistical Inference}
\label{sec:StatisticalInference}

The solar neutrino signal is searched using an unbinned extended likelihood fit~\cite{alea} performed jointly over the SR0 and SR1b datasets in the reconstructed recoil energy space between $[1,\,140]\,\si{keV}$.
For each science run $r$, the likelihood is constructed as
\begin{multline}
\mathcal{L}_r(\mu_{s,r}, \boldsymbol{\mu}_{b,r}, \boldsymbol{\theta})
= \mathrm{Poiss}(N_r \mid \mu_{\mathrm{tot},r}) \\
\times \prod_{i=1}^{N_r}
\left(
\frac{\mu_{s,r}}{\mu_{\mathrm{tot},r}} f_{s,r}(E_i,\boldsymbol{\theta})
+
\sum_j
\frac{\mu_{b_j,r}}{\mu_{\mathrm{tot},r}} f_{b_j,r}(E_i,\boldsymbol{\theta})
\right) \\
\times
\prod_m C_{\mu_m}(\mu_{b_m,r})
\prod_n C_{\theta_n}(\theta_n).
\label{Eqn:likelihood}
\end{multline}
with $\mu_{\mathrm{tot},r} \equiv \mu_{s,r} + \sum_j \mu_{b_j,r}$. Here $N_r$ is the number of observed events, $\mu_{s,r}$ and $\boldsymbol{\mu}_{b,r}$ are the rate parameters describing the expected signal and background event counts, and $E_i$ is the energy of the $i$-th event. The functions $f_{s,r}$ and $f_{b_j,r}$ are the normalized signal and background spectral templates. The non-rate systematic nuisance parameters $\boldsymbol{\theta}$ describe detector response (e.g., fiducial volume, efficiency, energy bias and resolution) and theoretical uncertainties in the spectral shapes of different isotope decays. The functions $C_{\mu_m}$ and $C_{\theta_n}$ denote the Gaussian constraints on the rate and non-rate systematic parameters, respectively. The combined likelihood for the joint fit is $\mathcal{L} = \prod_r \mathcal{L}_r$, from which the signal strength and its uncertainty are extracted via a scan of the profile likelihood ratio
\begin{equation}
q(\mu) = -2\ln\frac{\mathcal{L}(\mu, \hat{\hat{\boldsymbol{\theta}}}(\mu))}{\mathcal{L}(\hat{\mu}, \hat{\boldsymbol{\theta}})},
\end{equation}
where $\mu$ is the expected solar neutrino signal event count, $\hat{\mu}$ and $\hat{\boldsymbol{\theta}}$ are the global maximum-likelihood estimates, and $\hat{\hat{\boldsymbol{\theta}}}(\mu)$ are the nuisance parameters that maximize the likelihood for a fixed $\mu$.

\begin{table}[tb]
    \setlength{\extrarowheight}{2pt}
	\centering
    \caption{Total uncertainty budget for best-fit solar neutrino signal event count. The contributions from the non-rate systematic parameters are listed, and their combined uncertainty is reported as the ``Systematic uncertainty''. The uncertainty from the profile-likelihood fit, including both the statistical uncertainty and the contributions from the rate parameters, is also reported, as ``Profile-likelihood uncertainty''. The $\sigma_{-}$ and $\sigma_{+}$ are lower and upper relative uncertainties, respectively. The contributions from theoretical uncertainties in the spectral shapes of the other isotopes and in the SSM solar neutrino flux component fractions are found to be negligible. }
    \label{Tab:postInferenceSystematics}
	\begin{tabular}{l |c c}
		\hline
		& $\sigma_{-}$ (\%) & $\sigma_{+}$ (\%) \\
		\hline
		Fiducial volume &  9.7 & 10  \\
		Efficiency &  9.0 & 7.7 \\
        Energy bias &  2.9 & 4.1 \\
		Energy resolution &  1.3 & 1.4 \\
    	$^{214}$Pb correlated uncertainty &  4.3 & 4.1 \\
		$^{214}$Pb spectral shape &  1.1 & 1.1\\
		$^{136}$Xe spectral shape &  1.0 & 1.0 \\
		\hline
		Systematic uncertainty & 14 & 14 \\
		Profile-likelihood uncertainty & 14 & 14 \\
		\hline
        Total &  20 & 20 \\
        \hline
	\end{tabular}
\end{table}

\begin{table*}[t]
    \setlength{\extrarowheight}{2pt}
    \centering
    \newcolumntype{C}{>{\centering\arraybackslash}m{2.3cm}}
    \newcolumntype{L}{>{\raggedright\arraybackslash}m{3.0cm}}
    \caption{Nominal and corresponding best-fit numbers of events for SR0 and SR1b, with exposures of 1.09 t$\cdot$y and 1.37 t$\cdot$y, respectively. ``Solar $\nu$'' represents the measured total solar neutrino signal. Nominal values with an uncertainty are constrained in the fit by a Gaussian prior, while ``--'' denotes components absent in that science run.}
    \begin{tabular}{L | CC  CC}
        \hline
        & \multicolumn{2}{c}{SR0} & \multicolumn{2}{c}{SR1b}\\
        & Nominal & Best-fit & Nominal & Best-fit\\
\hline
 Solar $\nu$ & 295 & $ 500 \pm 70 $ & 387 & $ 660 \pm 90 $\\
 $^{124}$Xe DEC & 270 & $ 343 \pm 22 $ & 351 & $ 445 \pm 28 $\\
\hline
 $^{214}$Pb & $ 740 \pm 40 $ & $ 740 \pm 30 $ & $ 538 \pm 24 $ & $ 537 \pm 23 $\\
 $^{212}$Pb & $ 68 \pm 13 $ & $ 68 \pm 12 $ & $ 80 \pm 10 $ & $ 80 \pm 10 $\\
 $^{85}$Kr & $ 130 \pm 40 $ & $ 130 \pm 30 $ & $ 390 \pm 170 $ & $ 330 \pm 110 $\\
 Materials & $ 160 \pm 40 $ & $ 150 \pm 30 $ & $ 137 \pm 17 $ & $ 136 \pm 17 $\\
 $^{3}$H-like & -- & -- & 233 & $ 92 \pm 18 $\\
 $^{136}$Xe & $ 1460 \pm 50 $ & $ 1430 \pm 40 $ & $ 1990 \pm 70 $ & $ 1950 \pm 60 $\\
 \hline
 $^{133}$Xe & 140 & $ 310 \pm 60 $ & 15937 & $ 16740 \pm 140 $\\
 $^{83\mathrm{m}}$Kr & \text{free} & $ 98 \pm 14 $ & \text{free} & $ 111 \pm 19 $\\
 $^{125}$I & -- & -- & 39 & $ 0 \pm 30 $\\
 Accidentals & $ 0.4 \pm 0.3 $ & $ 0.5 \pm 0.3 $ & $ 1.3 \pm 0.1 $ & $ 1.3 \pm 0.1 $\\
\hline
    \end{tabular}
    \label{tab:expectation_values}
\end{table*}

The parameters in Eq.~(\ref{Eqn:likelihood}) fall into two categories that are treated differently in the inference. The first comprises the rate parameters $\boldsymbol{\mu}_b$, which scale the amplitude of the signal and background templates, providing the expected signal and background event counts. Those for which an external estimate is available (see Sec.~\ref{sec:constrainingTheBackgrounds}) enter the fit with Gaussian priors $C_{\mu_m}$; their nominal values are listed in the ``Nominal'' column of Table~\ref{tab:expectation_values}. The second category comprises the non-rate systematic parameters $\boldsymbol{\theta}$ listed in Table~\ref{Tab:postInferenceSystematics}. Several of these parameters, such as the fiducial volume and efficiency, are sufficiently degenerate with the rate parameters that including even a single one in the profile-likelihood maximization prevents the fit from converging. The effects of these systematic uncertainties are therefore estimated separately after the fit by varying each parameter individually by $\pm 1\sigma$, recording the resulting shift in the best-fit solar neutrino rate, and combining the individual contributions in quadrature to obtain the total systematic uncertainty $\sigma_{\rm sys}$. This procedure yields a $\sigma_{\rm sys}$ that is equivalent to a fully profiled fit for parameters degenerate with the rate multipliers (FV and efficiency), and conservative for those with additional shape information (energy scales and spectrum shapes). The total uncertainty is obtained by convolving the statistical-only profile likelihood with a Gaussian of the width $\sigma_{\mathrm{sys}}$, similar to the procedure used in the Borexino CNO measurement~\cite{Borexino_CNO_2023}.

In the joint likelihood fit of SR0 and SR1b, the solar neutrino, the $^{136}$Xe, and the $^{124}$Xe DEC (see Appendix~\ref{app:dec}) rate parameters are shared between the two datasets; all other background rate parameters are fitted independently for each science run. A special treatment is applied to the $^{214}$Pb component: its rate is fitted independently in SR0 and SR1b, each constrained by a Gaussian prior using only the \textit{uncorrelated} uncertainty from Table~\ref{Tab:RnResults}. The \textit{correlated} uncertainty between the two SRs, arising from the shared proportionality factor in the fit (see Sec.~\ref{sec:constrainingTheBackgrounds}), is not included in the priors but is propagated after the fit as an additional systematic uncertainty.

\noindent
\section{Results and Discussion}
\label{sec:results_discussion}

\begin{figure*}[t]
	\centering
	\includegraphics[width=\textwidth]{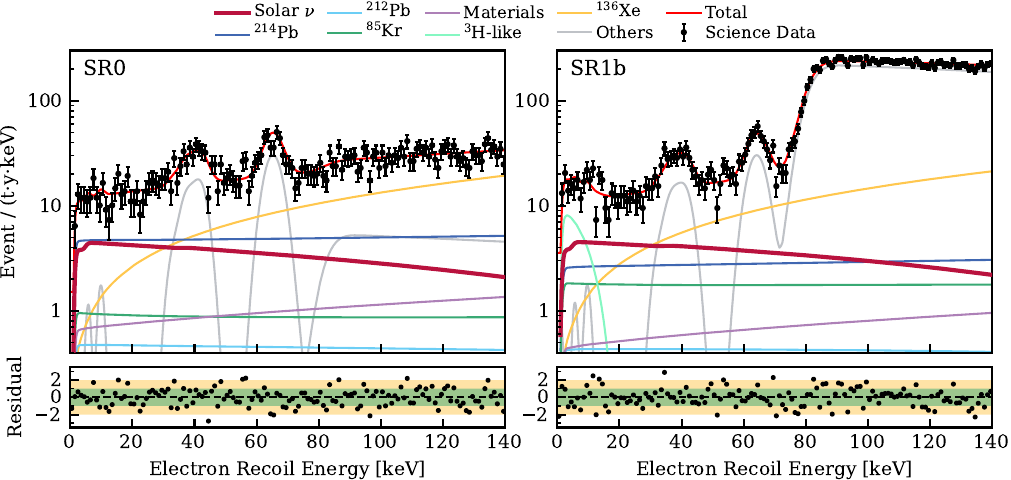}
	\caption{Combined fit to the SR0 (left) and SR1b (right) data within the $[1,\,140]\,\si{keV}$ ROI. Black points show the science data with statistical uncertainties; the thin red curves show the total best-fit model. Colored curves show the individual signal and background components as indicated in the legend. 
    The gray curve combines all the components covered in ``Other background'' in Sec.~\ref{sec:constrainingTheBackgrounds}; each of these components ($^{124}$Xe, $^{133}$Xe, $^{125}$I, $^{83m}$Kr, and Accidentals) is shown separately in Fig.~\ref{Fig:7_sr01_combined_fit_sr01_spectrum_C14}. The increase in event rate above around \qty{80}{keV} is contributed by $^{133}$Xe induced by neutron activation (see Appendix~\ref{app:neutronActivation}). The bottom panels show data-to-model residuals normalized by the statistical uncertainties of the data.}
	\label{Fig:7_sr01_combined_fit_sr01_spectrum}
\end{figure*}

\begin{figure}[tb]
	\centering
	\includegraphics[width=0.92\columnwidth]{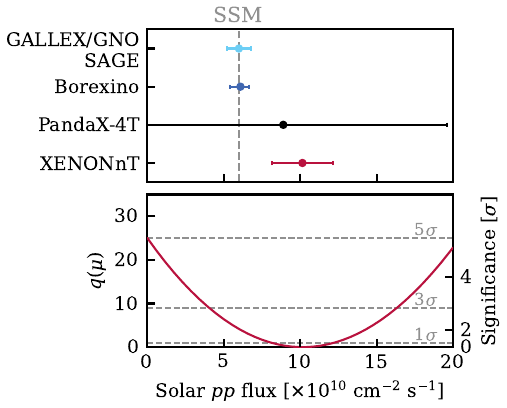}
	\caption{Solar $pp$ flux estimated from the inference results. Top: The best-fit result of this work is shown as the solid red point. Also plotted are results from combining GALLEX/GNO and SAGE~\cite{GALLIUM}, Borexino~\cite{Borexino}, and PandaX-4T~\cite{PandaX_solarpp}. The prediction from the B16-GS98 solar model is indicated by the vertical gray dashed line~\cite{SSM}. Bottom: Log likelihood ratio (Eq.~\ref{Eqn:likelihood}) as a function of solar $pp$ flux. Dashed lines indicate different significance levels.}
	\label{Fig:9_flux_scaled_only_pp_with_c14}
\end{figure}

The best fit of the energy spectrum with the solar neutrino signal model (FEA-stepping) is shown in Fig.~\ref{Fig:7_sr01_combined_fit_sr01_spectrum}, with the results summarized in Table~\ref{tab:expectation_values}. The goodness-of-fit is evaluated using the Pearson $\chi^2$ statistic, with the corresponding $p$-value of {\combinedpvalue} obtained from Monte Carlo simulations. The uncertainty from the best-fit profile-likelihood of the solar neutrino signal, which contains both the statistical uncertainty and the uncertainty of the rate parameters, is found to be \qty{14}{\percent}. The corresponding uncertainties for the non-rate systematic parameters that are itemized in Table~\ref{Tab:postInferenceSystematics} give a combined uncertainty of another \qty{14}{\percent}. None of the associated refits for the non-rate systematic parameters exhibits significant goodness-of-fit degradation. Accounting for the total uncertainty, the null hypothesis of no solar neutrino interactions is rejected with a statistical significance of {\solarnusignificancnominalcombined}.
Scaling the result with the solar $pp$ component fraction (Sec.~\ref{sec:solarpp_signal}), we obtain a solar $pp$ flux of {\solarppfluxresultnominalcombined}, consistent with the Borexino measurement~\cite{Borexino} within {\solarppdeviationnominalcombined}. 
The result is compared with the measurements from other experiments as well as the prediction from the B16-GS98 SSM\,\cite{SSM} in Fig.~\ref{Fig:9_flux_scaled_only_pp_with_c14}.
In addition, the result of the alternative RRPA signal model is reported in Appendix~\ref{app:rrpa}, and the measurement of the $^{124}$Xe half-life is reported in Appendix~\ref{app:dec}.

To test the larger but statistically consistent solar $pp$ neutrino flux against potential mismodeling, a series of cross-checks was performed: across different recoil energy intervals, within an inner \SI{2}{tonne} fiducial volume, projected to the $[140,\,200]\,\si{keV}$ region, and time dependence of the fit result by further splitting science runs into multiple periods, including the effect of annual modulation of the signal.
The cross-checks confirm that the result is robust and that the discovery power is mainly driven by the $[1,\,60]\,\si{keV}$ range. Potential contributions from $^{39}$Ar~(see Appendix\,\ref{app:ar39_background}), $^{218}$Po $\beta$-decays, and $^{134}$Xe two-neutrino double-$\beta$ decay~\cite{Xe136_HSD_and_Xe134} are also found to be negligible. 

Another potential background candidate is \isotope[14]{C}, with a Q-value of \SI{156.5}{keV} and half-life of \SI{5700}{y}~\cite{NuDataC14}. Although \isotope[14]{C} is not expected to be present in the detector, it is further investigated because its $\beta$ spectrum closely resembles that of the solar neutrinos and its concentration has never been measured at a sensitivity relevant to the low-background LXe environments of XENONnT. Adding \isotope[14]{C} to the background at its highest conceivable concentration results in a lower significance of {\solarnusignificanccarbon} and increases the consistency with Borexino to within {\solarppdeviationcarboncombined}. Further details are summarized in Appendix~\ref{app:c14}.

\section{Conclusion and Outlook}
\label{sec:conclusion_outlook}

Building on the newly established role of LXe TPCs as the most compact solar-neutrino detectors, we performed a measurement of solar neutrinos using data from the first two science runs of XENONnT with an exposure of {\exposure}.
Owing to the low energy threshold, the detector is sensitive to neutrino energies down to {\neutrinoenergylow}, which is one order of magnitude lower than in the previous measurement by Borexino \cite{Borexino}, and the lowest minimum detectable neutrino energy achieved to date.
Dedicated analysis methods were developed to improve the constraints on detector-related systematics and on key backgrounds, particularly from $^{214}$Pb.
We observe the scattering of solar neutrinos on xenon atomic electrons with a statistical significance of {\solarnusignificancnominalcombined}. The solar neutrino flux is dominated by neutrinos released in the $pp$ reaction, for which we derive a flux of {\solarppfluxresultnominalcombined}. This value is consistent with {\solarppdeviationnominalcombined} to the corresponding value reported by Borexino.
This measurement demonstrates the potential of XENONnT as a multipurpose liquid-xenon detector. 

Achieving a precision measurement of the solar $pp$ neutrino flux in LXe TPCs will require further reductions of and tighter constraints on the relevant backgrounds. These include not only the dominant contributions from $^{214}$Pb and $^{85}$Kr, but also those from $^{212}$Pb, $\gamma$-rays from detector materials, and potentially $^{14}$C. Improved modeling of the signal detection efficiency and reduced uncertainties in the fiducial volume will also be essential.

Following a recent upgrade campaign, XENONnT will continue to collect more data, allowing for measurements of solar neutrinos and their properties with increased exposure and a better understanding of detector systematics. Furthermore, next-generation experiments such as XLZD~\cite{XLZD}, which is projected to constrain the solar $pp$ neutrino flux at the sub-percent level, will enable precision tests of electroweak and neutrino physics, as well as studies for astrophysics such as solar metallicity and supernova neutrinos. 
They will also provide sensitivity to a broad range of Beyond Standard Model physics, including neutrinoless double-beta decay of $^{136}$Xe and dark matter interacting via the electronic-recoil channel in a multipurpose detector.

\noindent
\section*{Acknowledgments}
We thank Cheng-Pang Liu, Evgeny Akhmedov, and Jenni Kotila for fruitful discussions. We gratefully acknowledge support from the National Science Foundation, Swiss National Science Foundation, German Ministry for Education and Research, Max Planck Gesellschaft, Deutsche Forschungsgemeinschaft, Helmholtz Association, Dutch Research Council (NWO), Fundacao para a Ciencia e Tecnologia, Weizmann Institute of Science, Binational Science Foundation, Région des Pays de la Loire, Knut and Alice Wallenberg Foundation, Kavli Foundation, JSPS Kakenhi, JST FOREST Program, and ERAN in Japan, Tsinghua University Initiative Scientific Research Program, National Natural Science Foundation of China, Ministry of Education of China, DIM-ACAV+ Région Ile-de-France, and Istituto Nazionale di Fisica Nucleare. This project has received funding/support from the European Union’s Horizon 2020 and Horizon Europe research and innovation programs under the Marie Skłodowska-Curie grant agreements No 860881-HIDDeN and No 101081465-AUFRANDE. We gratefully acknowledge support for providing computing and data-processing resources of the Open Science Pool and the European Grid Initiative, at the following computing centers: the CNRS/IN2P3 (Lyon - France), the Dutch national e-infrastructure with the support of SURF Cooperative, the Nikhef Data-Processing Facility (Amsterdam - Netherlands), the INFN-CNAF (Bologna - Italy), the San Diego Supercomputer Center (San Diego - USA) and the Enrico Fermi Institute (Chicago - USA). We acknowledge the support of the Research Computing Center (RCC) at The University of Chicago for providing computing resources for data analysis. We thank the INFN Laboratori Nazionali del Gran Sasso for hosting and supporting the XENON project.

\vspace{2\baselineskip}

\textit{Note Added} -- We note the recent preprint from the PandaX-4T Collaboration reporting on a similar result of elastic solar neutrino-electron scattering with a significance of 2.2\,$\sigma$\,\cite{PANDAX_Solarpp2}.

\appendix
\noindent
\section{DETAILS ON EFFICIENCY EVALUATIONS}
\label{sec:appendix_efficiency}

The detection of an energy deposition in the TPC proceeds through a sequence of stages~\cite{XENONnT_SignalReconstruction}: individual PMT signals are first grouped and identified as S1 and S2 peaks, then paired into events using the open source reconstruction framework \texttt{straxen}~\cite{straxen}. The reconstructed events are then filtered by analysis-level selection criteria, and finally restricted to a central fiducial region of the target. 
Each of these stages, including the peak reconstruction, the event building, data quality selections, and fiducial volume definition, can introduce an efficiency loss, and their product defines the total detection efficiency in the reconstructed energy space. Mismodeling this efficiency, either in its central value or in its uncertainty, directly biases the inferred signal rate and degrades the discovery sensitivity, making careful evaluation a critical ingredient of the $pp$ neutrino flux measurement. The primary calibration source for these efficiency evaluations is $^{220}$Rn data, which provides a high-statistics electronic-recoil sample dominated by $\beta$ decays of $^{212}$Pb, a daughter isotope that distributes homogeneously throughout the LXe target. Additional datasets and simulations are used as supporting references, as discussed in the corresponding subsections.

\subsection{Peak reconstruction efficiency}
\label{app:PeakReconstructionEfficiency}
Peaks are formed by grouping single PMT signals with a minimal gap of less than \qty{700}{ns} to other adjacent PMT signals. These groups are then classified into either S1 peaks, S2 peaks or unknown based on the rise time of their summed waveform and the light distribution over the two PMT arrays. Losses in the peak reconstruction efficiency are dominated by losses in the S1 peak building and classification. To ensure that S1 signals arise from genuine scintillation signals and not an accidental pile-up of single-photon signals, each S1 is required to have at least three distinct PMT channels contributing within a $\pm 50\,\mathrm{ns}$ time window around the peak's amplitude. This 3-fold requirement sets the energy threshold of the detector at around \qty{1}{keV} for electronic recoils, which represent the lower limit of the ROI in the presented search. More details about the peak reconstruction and classification can be found in~\cite{XENONnT_SignalReconstruction}.   

\subsection{Event reconstruction efficiency}
\label{app:EventReconstructionEfficiency}

The event-building efficiency (EBE) quantifies the probability that an energy deposition is successfully reconstructed as an event with a correctly paired S1 and S2. Two effects can reduce the EBE: detector artifacts, such as single-electron (SE) peaks following large S2 signals that obscure the S1/S2 pairing, and dedicated selections that reject accidental coincidences (ACs), in which uncorrelated S1 and S2 signals are mistakenly paired into a single event~\cite{XENONnT_SR0_WIMP}. While ACs are negligible relative to the radioactive backgrounds in science data, they are non-trivial in high-rate calibration data, where pile-up of uncorrelated signals is more frequent. The EBE is therefore evaluated by injecting peak waveforms drawn from data back into the data stream at random times, and tracking which injected events fail the event-building procedure~\cite{saltax, axidence}. The resulting loss is quantified in the S1/S2 parameter space and then translated to the reconstructed energy space using the $^{220}$Rn calibration data.

Applying this procedure to the $^{220}$Rn calibration data is particularly challenging because the short half-life of $^{220}$Rn causes strong rate variations across the calibration period. We find that the shape of the EBE energy dependence remains consistent within each science run, while its overall normalization varies with rate; the EBE is therefore corrected on a run-by-run basis. The associated uncertainties, including potential simulation mismodeling and the limited statistics of individual runs, are propagated through to the final efficiency~\cite{appletree}. To be complete, we independently evaluated the EBE for the high-energy events from the $\alpha$ decays that are used in $^{214}$Pb background estimation (Sec.~\ref{subsec:radonAnalysis}), taking into account changes during calibration. The EBE for the $\alpha$ is found to be energy independent.

\subsection{Acceptance of data selection criteria}

Additional data quality selections based on the signal topology and shape of S1 and S2 signals ensure that only well-reconstructed events are considered in the inference. The acceptance of each selection criterion is evaluated using the $(n{-}1)$ data-driven method, in which the acceptance of one selection criterion is measured on a sample to which all other selection criteria have already been applied~\cite{XENONnT_SignalReconstruction}. The primary reference samples are the $^{220}$Rn calibration for SR0 and the SR1a science data for SR1b, both providing high-statistics ER events for the acceptance calculation. The statistical uncertainty on the acceptance in each energy bin is estimated via both the Clopper-Pearson exact interval and a Bayesian credible interval with a uniform prior. As a conservative treatment, the larger of the two is taken in each bin.

To account for potential mismodeling of the primary measurement (i.e., $^{220}$Rn for SR0 and SR1a data for SR1b), each acceptance is cross-checked against supporting datasets, such as $^{222}$Rn calibration data and \texttt{Geant4}-based waveform simulations~\cite{Geant4, fuse}. A systematic uncertainty $\sigma_{\mathrm{sys}}$ is then assigned to enforce statistical consistency between the primary and supporting measurements, defined by
\begin{equation}
\chi^2 = \frac{(A_{\mathrm{supp}} - A_{\mathrm{prim}})^2}{\sigma_{\mathrm{prim}}^2 + \sigma_{\mathrm{sys}}^2 + \sigma_{\mathrm{supp}}^2} \le 1,
\end{equation}
where $A_{\mathrm{prim}}$ and $A_{\mathrm{supp}}$ are the acceptances measured on the primary and supporting datasets. The value of $\sigma_{\mathrm{sys}}$ is inflated until this condition is met, and contributes to the upper (lower) uncertainty when the supporting dataset yields a higher (lower) acceptance.

For selections whose acceptance depends on the local rate of nearby peaks around each event, a further correction is applied. Because the ambient rate is significantly higher in calibration than in science data, the calibration-derived acceptance would be biased. To quantify this effect, simulated peaks are injected into both calibration and science data~\cite{saltax, axidence}; the median acceptance is taken as the nominal value and half the difference as its uncertainty. The total impact on the data selection acceptance is approximately \qty{2}{\percent}.

The accumulated acceptance across all selection criteria is modeled as a single cubic polynomial spanning $[0,\,200]\,\si{keV}$, extending beyond the ROI. This smooth functional form captures the overall energy dependence of the acceptance while smoothing out statistical fluctuations in individual bins, which is well-suited to the relatively flat spectral features of both the signal and background in the solar neutrino analysis. Spatial variations of the acceptance are incorporated as a systematic uncertainty applied after the fit. No significant time variation is observed in the science data, and therefore, no additional time-dependent systematic is assigned. 

\subsection{Fiducial volume}
\label{sec:appendix_fdv}

To facilitate a three-dimensional position reconstruction of the interaction vertex with a $\mathcal{O}(\qty{5}{\milli \meter})$ detector resolution, the depth ($z$) of the interaction is measured through the drift time along the height of the cylinder, while the S2 signal pattern measured by the top PMT  array contains information about the position parallel to the cylinder base ($x$ and $y$). This allows the fiducial volume (FV) selection of the central, cleanest region of the LXe target. It is optimized in the $(R^2, Z)$ coordinate space, where spatial bins are ranked by their signal-to-background ratio ($s/b$) based on the full-chain simulation with \texttt{Geant4}~\cite{Geant4} and \texttt{fuse}~\cite{fuse}. Here, $s$ denotes the homogeneously distributed signal events expected inside the TPC, and $b$ denotes the material $\gamma$ events that are concentrated near the TPC surface. The optimal region maximizes the figure of merit,
\begin{equation}
\mathrm{FOM} = \sqrt{2\left[(s+b)\ln\left(1+\frac{s}{b}\right) - s\right]},
\end{equation}
corresponding to the Asimov significance~\cite{FOM}. The resulting statistically optimal contour is simplified to a polygon in the $(R^2, Z)$ space that is conservatively inscribed within it. The optimization is validated against $^{83\mathrm{m}}$Kr calibration data and SR1a data, where the FV boundary is shown to exclude the regions in which the event rate rises steeply toward the TPC edge.

A volumetric uncertainty of less than \qty{0.1}{\percent} is obtained from simulation by comparing the number of events within the FV between their ground truth and reconstructed positions. To remain conservative against potential mismodeling of the field-distortion correction and the charge-insensitive region at the TPC edges, an additional \qty{3}{\percent} systematic uncertainty is assigned to the FV~\cite{ToschiThesis}. The LXe density was estimated based on the monitored temperature and pressure readings of the TPC and found to be \qty{2861.5\pm3.1}{\kilo \gram \per \cubic \meter} and \qty{2859.5\pm3.3}{\kilo \gram \per \cubic \meter} for SR0 and SR1, respectively. This results in a fiducial mass of $(4.13 \pm 0.12)$~tonnes for SR0 and $(4.24 \pm 0.13)$~tonnes for SR1.

\section{DETAILS ON BACKGROUND EVALUATIONS}
\label{app:details_background_eval}

\subsection{Argon-39}
\label{app:ar39_background}

The $\beta$-emitter $^{39}$Ar has a half-life of 269\,y, an endpoint of 565~keV, and an abundance of $7.7~\times~10^{-16}$~mol/mol in $^{\rm nat}$Ar~\cite{adhikari2023precision}. Its contribution to the ER background is suppressed by an online cryogenic distillation campaign performed prior to the start of SR0, lasting for 47 days~\cite{kryptonDistillation}. Although originally developed for krypton removal, this system is even more efficient for argon, which is over four times more volatile than krypton and can be extracted with a significantly shorter time constant~\cite{kryptonDistillation, argonDistillation}. The effective removal time constant of the system for argon was measured to be $\tau_{\rm dist} = 3.8$~d using the $^{37}$Ar rate evolution after the SR0 $^{37}$Ar calibration campaign. Applying this time constant to the 47-day campaign, and starting from the initial $^{\rm nat}$Ar/Xe concentration of $<5\times 10^{-6}$ mol/mol~\cite{XENONnT_SR0_lowER}, the residual concentration is reduced by a factor of about $4 \times 10^{-6}$ to $<2\times 10^{-11}$ mol/mol, corresponding to a $^{39}$Ar rate below $0.1$~event$/$(t$\cdot$y) within the ROI. A similar argument can be made for the SR1b data. This contribution is therefore negligible.

\subsection{Radon $\alpha$ analysis}
\label{app:rn_alpha_analysis}

The $\beta$ decays of the radon progeny \isotope[214]{Pb} and \isotope[212]{Pb} constitute the dominant background sources for the measurement of solar neutrinos. Therefore, their rate must be determined independently via the $\alpha$-decaying isotopes present in their decay chains (Eq.~(\ref{eq:rn222_chain}) and (\ref{eq:rn220_chain})).

Given the distinct signal topology of mono-energetic $\alpha$ decays, the energy reconstruction using the S1 and S2 produces well-isolated Gaussian peaks \,\cite{radonRemoval_level}, from which the $\alpha$ rate can be estimated using the event counts within the peak energy range. However, for isotopes such as \isotope[214]{Po} (\isotope[212]{Po}), following the decay chains from Eq.~(\ref{eq:rn222_chain}) (Eq.~(\ref{eq:rn220_chain})), a complication arises due to their short half-life and their preceding $\beta$ decays of \isotope[214]{Bi} (\isotope[212]{Bi}), which decay into an excited nuclear state and promptly emit $\gamma$'s from de-excitation. As the $\gamma$ disrupts the S2 peak reconstruction~\cite{XENONnT_Bi214Spec}, estimating the \isotope[214]{Po} $\alpha$ rate requires a specialized event processing relying only on the S1 signal\,\cite{radonRemoval}. This involves independent signal corrections, selections, and efficiencies, as well as position reconstruction, to obtain mono-energetic energy peaks with accurate counts. The efficiency and fiducial volume are much more weakly constrained when evaluated only with S1 signals. To improve the constraint, they are instead determined by referencing each event to the analysis using both S1 and S2 signals. The \isotope[222]{Rn} and \isotope[218]{Po} events are used as the reference because, in the absence of analysis biases, both $\alpha$ analyses should yield identical results. Finally, the decay rate of each $\alpha$-decay isotope is then determined by the counts falling within a $\pm 3\sigma$ range of the peak's mean position over 24-hour intervals. 

\begin{table}[t]
    \setlength{\extrarowheight}{2pt}
	\centering
    \caption{Estimation of the $\alpha$ rate and plate-out ratios. For the $^{212}$Pb/\isotope[212]{Bi}\isotope[212]{Po} ratios, the value is taken as the average of the $^{212}$Pb/$^{212}$Po and $^{212}$Pb/$^{212}$Bi after correcting for their respective branching ratios. }
	\begin{tabular}{c|cc}
	\hline
        \hline
	    & SR0 & SR1b \\
        \hline
        $^{222}$Rn activity (\qty{}{\micro \becquerel \per \kilo \gram}) & 2.044 $\pm$ 0.009 & 1.141 $\pm$ 0.006 \\
        $^{214}$Pb/$^{222}$Rn ratio & 0.68 $\pm$ 0.03 & 0.67 $\pm$ 0.03 \\
        $^{214}$Po/$^{222}$Rn ratio & 0.425 $\pm$ 0.004 & 0.411 $\pm$ 0.005 \\
        \hline
        $^{212}$Po activity (\qty{}{\micro \becquerel \per \kilo \gram}) & 0.0156 $\pm$ 0.0011 & 0.0168 $\pm$ 0.0011 \\
        $^{212}$Pb/\isotope[212]{Bi}\isotope[212]{Po} ratio & 2.4 $\pm$ 0.2 & 2.05 $\pm$ 0.02 \\
        \hline
        \hline
	\end{tabular}
	\label{Tab:RnAppendix}
\end{table}

For the \isotope[214]{Pb} rate estimation, the analysis is based on the \isotope[222]{Rn} calibration performed only in SR1b. While detector conditions, such as pressure, temperature, liquid level, and electric fields, are found to be stable within each of the two SRs, there are minor differences between them as reported in~\cite{XENONnT_SR1_WIMP, XENONnT_B8_SR1}.
The influence of these differences on the activity ratio between \isotope[222]{Rn} and \isotope[214]{Pb} is estimated using the \isotope[214]{Pb} daughter isotope \isotope[214]{Po}.  This is achieved by comparing the $\alpha$-rate ratio of \isotope[214]{Po} to \isotope[222]{Rn} during the calibration period with the corresponding ratios measured in SR0 and SR1b. A systematic difference of \SI{4.8}{\percent} is observed and used to correct the \isotope[214]{Pb} rate estimate, with half of this difference being assigned as an uncertainty. Furthermore, the spatial non-uniformity of the $\alpha$ event rate between the SRs was studied and found to have a negligible impact. 

\begin{figure*}[t]
	\centering
	\includegraphics[width=\textwidth]{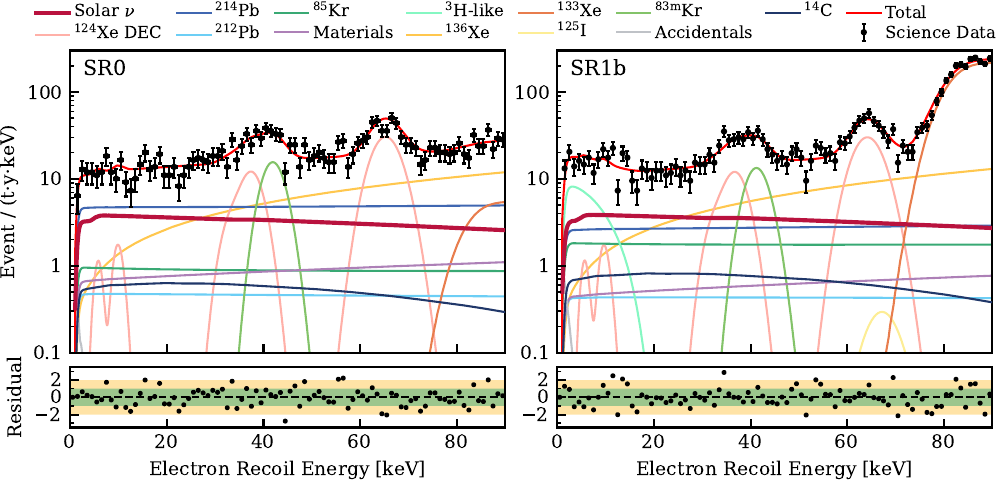}
	\caption{Combined fit analogous to Fig.~\ref{Fig:7_sr01_combined_fit_sr01_spectrum}, with the addition of the hypothetical $^{14}$C background component and the ``Others'' background separated into individual components. The range is set to $[0,\,90]\,\si{keV}$ to provide the reader with more information about the highest sensitive region, while no new pattern above \SI{90}{\keV} is visible.}
	\label{Fig:7_sr01_combined_fit_sr01_spectrum_C14}
\end{figure*}

As presented in Sec.~\ref{subsec:radonAnalysis}, the \isotope[212]{Pb} rate estimation relies on the $\alpha$ decays of \isotope[212]{Bi} and \isotope[212]{Po}, both of which are decay daughters of \isotope[212]{Pb}, as shown in Eq.~(\ref{eq:rn220_chain}). Because the \isotope[220]{Rn} calibration procedure uses several short-time source injections, the \isotope[212]{Pb} progeny does not reach its equilibrium activity during the period. This requires a more thorough modeling via a system of coupled differential Bateman equations\,\cite{Bateman:1910} describing each isotope's time-dependent activity. The model accounts for the instant source introduction, as well as the decay, removal, and empirical recovery effect for each isotope. This model is solved numerically using difference equations and matched simultaneously to the observed $^{212}$Po $\alpha$ rate and $^{212}$Pb rate in a $[10,\,70]\,\si{keV}$ window using a Markov Chain Monte Carlo method. The activity ratio between \isotope[212]{Pb} and \isotope[212]{Po} is then obtained from the steady-state solution of this model. The systematic uncertainties of the \isotope[212]{Pb} rate are estimated by comparing predictions derived using \isotope[212]{Po} and \isotope[212]{Bi} rates. The corresponding results of the activity ratio are collectively provided in Table~\ref{Tab:RnAppendix}.

\subsection{Neutron-activated Xe isotopes}
\label{app:neutronActivation}

The $^{133}$Xe $\beta$ decay, with a half-life of \SI{5.247}{d}~\cite{NuDataXe133}, is the primary time-dependent background induced by neutron activation from AmBe calibration performed in SR1b.
While the decay has a continuous spectrum, it is always followed by the de-excitation of an \SI{81.0}{\keV} excited state, contributing only above this energy level, as shown in Fig.~\ref{Fig:7_sr01_combined_fit_sr01_spectrum}. We modeled the time evolution by fitting the total count within the energy range of $[70,\,140]\,\si{keV}$ with exponential functions following AmBe injections. The fit gives a half-life of $(5.94 \pm 0.16)\,\mathrm{d}$, slightly longer than the literature value, which is likely due to the influence of other components within this wide energy window. On the other hand, for SR0, potential neutron activation of $^{133}$Xe can come from outside the TPC, such as in the cryogenic and distillation systems that are not shielded by the \qty{700}{tonne} water tank. Our estimate, based on the environmental neutron rate, yields a nominal rate of $(142 \pm 20)$~events/(t$\cdot$y) within the ROI. Since $^{133}$Xe lies largely outside the signal-dominated region and a dedicated Monte Carlo study shows that it has a percentage-level impact on projected sensitivity, we conservatively set the estimations as the nominal rates, while leaving the rate as a free parameter in the inference.

Another neutron-activated background is $^{125}$Xe, which decays into $^{125}$I with a half-life of \SI{16.9}{\hour}~\cite{NuDataXe125}. The subsequent electron capture of $^{125}$I into the excited state has a half-life of \SI{59.4}{\day}. The de-excitation of \SI{35.5}{\keV}~\cite{NuDataXe125} is again promptly followed by internal conversion and Auger cascade, producing a cumulative energy of \SI{67.3}{\keV} with a branching ratio of \qty{80.1}{\percent}~\cite{XENON1T_DEC_first}. We fit the $^{125}$Xe \SI{276.6}{\keV} emission line with an exponential time-evolution, and then estimate the $^{125}$I rate according to the Bateman equation. Although the estimate suggests a negligible contribution, as the uncertainty of the estimate is large and difficult to validate, we set the nominal value of the $^{125}$I rate according to the estimate but leave the parameter free. Fig. \ref{Fig:7_sr01_combined_fit_sr01_spectrum_C14} shows the contribution of the individual background components for the best solar neutrino fit. 

\section{SOLAR NEUTRINO RESULT WITH THE RRPA SIGNAL MODEL}
\label{app:rrpa}
 
For the FEA-stepping model as discussed in Sec.~\ref{sec:solarpp_signal}, atomic effects such as relativistic correction and electron–electron correlation are not accounted for. A cross-section model based on the Relativistic Random Phase Approximation (RRPA) has been introduced as a refinement~\cite{RRPAsignalModel}. The RRPA model predicts a cross section approximately \qty{25}{\percent} smaller than the FEA-stepping model, resulting in a reduced expected rate. However, the current RRPA calculations are evaluated up to $E \leq 32$\,keV, while our ROI extends to $140$\,keV. Given the reasoning that the RRPA cross-section converges to FEA at higher $E_r$, as communicated with the authors of Ref.~\cite{RRPAsignalModel}, we approximated the remaining $[32,\,140]\,\si{keV}$ by scaling down the FEA model to match the RRPA value at \SI{32}{keV}. Considering this approximated RRPA signal model, we obtain a statistical significance of {\solarppsignificancrrpa} and a solar $pp$ flux of {\solarppfluxresultrrpacombined}.

\section{XENON-124 DEC HALF-LIFE MEASUREMENT}
\label{app:dec}
In this study, $^{124}$Xe double-electron capture (DEC) is also treated as a signal channel, providing an independent validation of the inference framework. The $^{124}$Xe spectrum model includes the dominant \qty{72.5}{\percent} KK-shell capture peak, as well as 30 subdominant capture shell-levels. As the $^{124}$Xe spectra are distinctly peak-shaped, their rates are left free with a fixed branching ratio, such that the derived half-life can be compared to the literature measurements and theoretical prediction~\cite{DEC_Theory}. The nominal rate value is set to the weighted average of the experimental half-life measurements at $1.07 \times 10^{22}$ y. Together with an RGA-measured $^{124}$Xe abundance of $(10.1 \pm 0.6) \times 10^{-4}$ in our xenon inventory, our inference reports a half-life of {\dechalflife}, consistent with the literature value within {\decdeviation}.

\section{INFERENCE RESULT WITH CARBON-14}
\label{app:c14}

Given the close resemblance of the \isotope[14]{C} $\beta$ spectrum to that of the solar neutrino signal, it has been considered as a hypothetical background contribution. Its natural isotopic abundance is \num{1.18e-12}~\cite{C14abundance, C14variation}. Cosmogenic $^{14}$C is produced primarily through cosmic-ray interactions in the upper atmosphere and is subsequently incorporated predominantly into CO$_2$. Although the purification systems are expected to reduce the concentration of carbon-bearing molecules, including CO$_2$, to a low residual level, we consider CO$_2$ to be the most plausible carrier of residual $^{14}$C in the xenon.

The gaseous purification system has a quoted upper bound for CO$_2$ removal at \SI{100}{ppt}~\cite{CO2_100ppt}, limited by the sensitivity of the measurement method. This bound is applicable given that additional CO$_2$ from outgassing of detector materials was found to be negligible, based on residual gas analyzer measurements taken during the vacuum phase prior to filling the TPC with xenon. Incorporating $^{14}$C as an additional background with a constraint of $[0,\,100]\,\si{ppt}$ in CO$_2$, which is particularly conservative given that XENONnT also operates a specialized liquid purification system~\cite{liquidPurification}, the solar $pp$ flux becomes {\solarppfluxresultcarboncombined}. The statistical significance and agreement with Borexino become {\solarnusignificanccarbon} and {\solarppdeviationcarboncombined}, respectively. The energy spectrum under the best-fit including the \isotope[14]{C} background is shown in Fig.~\ref{Fig:7_sr01_combined_fit_sr01_spectrum_C14}. Even in this conservative estimate the$^{14}C$ contribution remains subdominant compared to other backgrounds.

\bibliography{bibliography}   

\end{document}